\documentclass[twocolumn]{63}
\usepackage{silence} 
\usepackage{natbib}
\usepackage{amsmath}
\usepackage{float}
\usepackage{times}
\usepackage{enumitem}
\usepackage{etoolbox}
\usepackage{graphicx}
\usepackage{amsmath}
\usepackage{bm}
\usepackage{graphics}
\usepackage{url}
\usepackage{epsfig}
\usepackage{wrapfig}
\usepackage{ulem}
\usepackage{verbatim}
\usepackage{float}
\usepackage{lastpage}
\usepackage{longtable}

\def\lsim{\hbox{\rlap{\raise 0.425ex\hbox{$<$}}\lower 0.65ex\hbox{$\sim$}}}
\def\gsim{\hbox{\rlap{\raise 0.425ex\hbox{$>$}}\lower 0.65ex\hbox{$\sim$}}}

\def\arcdeg{\mbox{$^\circ$}}

\shorttitle{GW190814 H$_0$}
\shortauthors{Vasylyev and Filippenko}

\begin{document}

\title{A Measurement of the Hubble Constant using Gravitational Waves from the Binary Merger GW190814}
\author{Sergiy~S.Vasylyev}
\affiliation{Department of Astronomy, University of California, Berkeley, CA 94720-3411, USA}
\affiliation{Steven Nelson Graduate Fellow}
\author{Alexei~V.~Filippenko}
\affiliation{Department of Astronomy, University of California, Berkeley, CA 94720-3411, USA}
\affiliation{Miller Senior Fellow, Miller Institute for Basic Research in Science, University of California, Berkeley, CA 94720, USA}

\begin{abstract}
We present a test of the statistical method introduced by Bernard F. Shutz in 1986 using only gravitational waves to infer the Hubble constant (H$_0$) from GW190814, the first high-probability neutron-star--black-hole (NS-BH) merger candidate detected by the Laser Interferometer Gravitational-wave Observatory (LIGO) and the Virgo interferometer. We apply a baseline test of this method to the binary neutron star (BNS) merger GW170817 and find H$_0 = 70^{+35.0}_{-18.0}$\,km\,s$^{-1}$\,Mpc$^{-1}$ (maximum {\it a posteriori} and 68.3\% highest density posterior interval) for a galaxy $B$-band luminosity threshold of $L_B \geq 0.001\,L_B^*$ with a correction for catalog incompleteness. Repeating the calculation for  GW190814, we obtain H$_0 = 67^{+41.0}_{-26.0}$\,km\,s$^{-1}$\,Mpc$^{-1}$ and  H$_0 = 71^{+34.0}_{-30.0}$\,km\,s$^{-1}$\,Mpc$^{-1}$ for  $L_B \geq 0.001\,L_B^*$ and $L_B \geq 0.626\,L_B^*$, respectively. Combining the posteriors for both events yields H$_0 = 70^{+29.0}_{-18.0}$\,km\,s$^{-1}$\,Mpc$^{-1}$, demonstrating the improvement on constraints when using multiple gravitational-wave events. We also confirm the results of other works that adopt this method, showing that increasing the $L_B$ threshold enhances the posterior structure and slightly shifts the distribution's peak to higher H$_0$ values.We repeat the joint inference using the low-spin PhenomPNRT \citep{the_ligo_scientific_collaboration_properties_2019} and the newly available combined \citep[SEOBNRv4PHM + IMRPhenomPv3HM;][]{abbott_gw190814_2020} posterior samples for GW170817 and GW190814, respectively, achieving a tighter constraint of H$_0 = 69^{+29.0}_{-14.0}$\,km\,s$^{-1}$\,Mpc$^{-1}$.
\end{abstract}



\section{Introduction}\label{s:intro}

Gravitational-wave (GW) and electromagnetic (EM) follow-up observations of black hole (BH) and neutron star (NS) mergers provide a novel method of probing dense astrophysical environments and enable a unique channel for cosmology. The August 2017 discovery of NS-NS merger GW170817 by the Laser Interferometer Gravitational-Wave Observatory (LIGO) and the Virgo interferometer, as well as the subsequent multimessenger observations by the coordination of thousands of astronomers, has brought forth a new era of astronomy \citep{ligo_scientific_collaboration_and_virgo_collaboration_gw170817_2017}. Two weeks after this discovery, the LIGO detectors in Hanford, Washington and Livingston, Louisiana, as well as the newly added Virgo detector in Italy, ceased operation for upgrades, marking the end of Observing Run 2 (O2). The analysis of GW170817 covered the entire EM spectrum, giving insights into the generation of short-duration gamma-ray bursts \citep[sGRBs;][]{goldstein_ordinary_2017}, general relativity in the strong-gravity regime \citep{the_ligo_scientific_collaboration_tests_2019}, the production of $r$-process elements \citep{kasen_origin_2017}, constraints on the NS equation of state \citep{annala_gravitational-wave_2018}, and an independent measurement of the Hubble constant H$_0$ \citep{abbott_gravitational-wave_2017}. 

Observing Run 3 (O3) began in late April of 2019, promising a higher rate of GW detections owing to sensitivity upgrades to the three detectors. The search volume increased by $\sim 100$\% from O2 to O3 according to Table 2 in \cite{kagra_collaborationligoscientificcollaborationandvirgocollaboration_prospects_2018}, thereby allowing for a significantly deeper survey. The end of O3 was initially planned for April 30, 2020 in order to conduct additional upgrades, but O3 was ended prematurely on March 27, 2020 because of the COVID-19 pandemic. In summary, O3 produced 36 BH-BH mergers, 1 NS-NS  merger, and the first-ever candidate NS-BH merger (GW190814) with at least 90\%  confidence. In this paper, we will focus on the candidate NS-BH merger GW190814, as its exquisite localization (see  \hyperref[sec:skymap]{Section 2.3}) makes it a prime object for the statistical inference of H$_0$ described below. Hereinafter, we will refer to GW190814 as a NS-BH merger, though we acknowledge that this event has not been explicitly shown to be a NS-BH; it likely contains compact objects with a mass range falling within the criterion for a NS-BH merger.\footnote{See LVC \href{https://emfollow.docs.ligo.org/userguide/content.html}{Public Alerts User Guide}'s criteria for GW event classification} 

\subsection{The Hubble Tension}

Recent measurements of H$_0$ reveal at least a 4.2$\sigma$ tension. The {\it Planck} satellite team infers H$_0 = 67.4 \pm 0.5$\,km\,s$^{-1}$\,Mpc$^{-1}$ using cosmic microwave background (CMB) data and assuming the standard $\Lambda$-CDM model is correct \citep{planck_collaboration_planck_2019}. The SH0ES (Supernovae, H$_0$, for the Equation of State of Dark Energy) team measured H$_0 = 74.03 \pm 1.42$\,km\,s$^{-1}$\,Mpc$^{-1}$ with an independent method, combining Cepheid-variable calibrations with luminosity distance measurements of Type Ia supernovae \citep{riess_milky_2018,riess_large_2019}. The SH0ES method is sensitive to how well the distance ladder is calibrated. Other recent studies have only enhanced the discrepancy between the two methods mentioned above, for a combined tension of $\sim 6\sigma$ \citep{riess_expansion_2020}. For example, the H0LiCOW (H$_0$ Lenses in
COSMOGRAIL’s Wellspring) and STRIDES (STRong-lensing Insights into Dark
Energy Survey) teams use measured time delays in the light curves of different images of single strongly lensed quasars to obtain H$_0 = 73.3 \pm 1.8$\,km\,s$^{-1}$\,Mpc$^{-1}$ and H$_0 = 74.2 \pm 1.4$\,km\,s$^{-1}$\,Mpc$^{-1}$, respectively, agreeing with the SH0ES measurement \citep{wong_h0licow_2019,shajib_strides_2020}. Other measurements that broadly agree with the SH0ES value include the tip of the red giant branch \citep[TRGB1, TRGB2;]{jang_tip_2017,hatt_carnegie-chicago_2018}, Mira variables \citep{huang_hubble_2020}, surface brightness fluctuations (SBF), and masers \citep{verde_tensions_2019}. On the other hand, measurements from big-bang nucleosynthesis and baryon acoustic oscillations \citep[BBN +  BAO;][]{cuceu_baryon_2019}, Wilkinson Microwave Anisotropy Probe (WMAP) CMB + BAO \citep{hinshaw_five-year_nodate}, Atacama Cosmology Telescope Polarization camera (ACTPol) + BAO \citep{louis_atacama_2017}, and the South Pole Telescope SZ camera (SPT-SZ) + BAO \citep{story_measurement_2013} favor the {\it Planck} result.

\subsection{The Standard-Siren Method}

The above tension between local (SH0ES, etc.) and early-universe ({\it Planck}) measurements of  H$_0$ may be the result of underlying systematic effects or astrophysical causes. A new independent method using gravitational-wave sources as ``standard sirens," first proposed by \cite{schutz_determining_1986}, could reconcile this discrepancy. This method uses the observed amplitude and the frequency of the gravitational waveform (from which a distance  is determined) together with the measured redshift of the source's host galaxy to infer H$_0$  \citep{holz_using_2005,nissanke_exploring_2010,vitale_measuring_2018,mortlock_unbiased_2019}. Standard sirens do not need a distance ladder and are thereby decoupled from systematic effects that may be introduced in methods relying on it. 

The true host galaxy can only be identified with an EM counterpart. GW170817 produced an optical transient powered by radioactive decay, called a kilonova \citep{kasen_origin_2017}. The source's host galaxy was identified as NGC 4993 at a luminosity distance of $40^{+8.0}_{-14.0}$\,Mpc \citep{coulter_swope_2017,soares-santos_electromagnetic_2017,valenti_discovery_2017,arcavi_optical_2017,tanvir_emergence_2017,lipunov_master_2017}. \cite{abbott_gravitational-wave_2017} presented the first result using the standard  siren method, estimating H$_0 = 70.0^{+12.0}_{-8.0}$\,km\,s$^{-1}$\,Mpc$^{-1}$.
\\

\subsection{The Galaxy-Catalog Method}
 BH-BH mergers are not expected to produce an optical transient, unlike NS-NS mergers. Some theoretical models suggest that NS-BH mergers may only produce a counterpart under certain physical conditions \citep{foucart_remnant_2018,foucart_numerical_2019,barbieri_electromagnetic_2020}. However, cosmological  inferences are still possible without the EM counterpart. 
 
 The LIGO-Virgo detectors produce a skymap that constrains
 the location of a GW source to specific patches in the sky. Using a
 statistical approach (the catalog method), one can consider
 every galaxy in the source's localization region as a potential host with some probability. Summing the probability assigned to each galaxy builds the full posterior on H$_0$.  \cite{chen_2_2018} show that an H$_0$
 inference from GW sources with optical counterparts will converge faster than for dark sirens. Although using a so-called ``dark siren"
 will not provide as precise a measurement from a single
 event compared to the case when the host galaxy is known,
 many more BH-BH mergers (dark sirens) are expected than
 NS-NS mergers \citep{baibhav_gravitational-wave_2019}, providing a useful validation test of the optical-counterpart method. Furthermore, combining several measurements will yield increasingly tighter constraints on $\text{H}_0$ \citep{chen_2_2018,nair_measuring_2018,feeney_prospects_2019}.
 
 The galaxy catalog (statistical) method was first tested on simulated data by \cite{del_pozzo_inference_2012}. More recently, \cite{fishbach_standard_2019} used this method to infer H$_0$ from
 GW170817 without relying on the EM counterpart. They obtain several estimates for H$_0$ using various luminosity cuts and weighting schemes to galaxies in the GLADE 2.3 catalog described in \hyperref[sec:catalog]{Section 2.2}. An estimate of H$_0$ from
 BH-BH merger GW170814 using a similar statistical method
 was recently obtained by the Dark Energy Survey (DES)
 Year-3 data team with a proprietary galaxy catalog (DES Y3; \citealt{abbott_dark_2018}). The DES team computed H$_0 = 75^{+40.0}_{-32.0}$ and H$_0 = 78^{+96.0}_{-24.0}$\,km\,s$^{-1}$\,Mpc$^{-1}$ for the uniform prior ranges [20,~140] and [10,~220]\,km\,s$^{-1}$\,Mpc$^{-1}$, respectively \citep{the_des_collaboration_first_2019}. The LIGO-Virgo Collaboration (LVC) combined high-probability BH-BH (dark sirens) from the O1 and O2 runs together with the GW170817 optical counterpart, yielding a joint value H$_0 = 68^{+14.0}_{-7.0}$ with a [20,~140]\,km\,s$^{-1}$\,Mpc$^{-1}$ flat-in-log H$_0$ prior defined in \hyperref[sec:priors]{Section 2.1} \citep{the_ligo_scientific_collaboration_gravitational-wave_2019}. This work also explored the effects of galaxy luminosity weighting on the H$_0$ posterior shape.
 \newpage
 \subsection{GW190814}
 On 2019-08-14, at 21:10:39 UT, LIGO Hanford, LIGO Livingston, and Virgo detected the GW event GW190814 with a false-alarm rate (FAR) of approximately 1 per $10^{25}$\,yr at a luminosity distance of $267 \pm 52$\,Mpc \citep{ligo_scientific_collaboration_ligovirgo_2019}.
 Although the NS-BH candidate GW190814 did not have an associated
 EM counterpart, we can still use the gravitational-wave data to produce meaningful results. Analysis by \cite{abbott_gw190814_2020} showed that the primary and secondary masses of GW190814 are $23^{+1.1}_{-1.0}\,M_{\odot}$ and $2.59^{+0.08}_{-0.09}\,M_{\odot}$, respectively. The secondary mass approaches the observational MassGap (3--5\,$M_{\odot}$), in which there is uncertainty regarding whether the object is the heaviest neutron star or the lightest black hole ever discovered. This should not affect our results, given our generous prior on the NS mass for the event. After initial submission of this paper, \cite{abbott_gw190814_2020} performed the statistical method for GW190814 also using the GLADE catalog to obtain H$_0 = 75^{+59.0}_{-13.0}$\,km\,s$^{-1}$\,Mpc$^{-1}$ with a flat H$_0$ prior on  [20,~140] \,km\,s$^{-1}$\,Mpc$^{-1}$ and using posterior samples.
 
 We follow the methodology presented by \cite{chen_2_2018}, \cite{gray_cosmological_2019}, and \cite{the_ligo_scientific_collaboration_gravitational-wave_2019} to obtain the
 H$_0$ posterior. With this paper we test the statistical method on a new GW-type candidate (NS-BH) and improve the accessibility of the gwcosmo \footnote{https://git.ligo.org/lscsoft/gwcosmo/-/tree/master} code. See the Appendix for a detailed discussion of the mathematics involved.

\section{Methods}\label{s:meth}
Using the publicly available gwcosmo code, we construct a posterior on the Hubble constant using only gravitational waves for the NS-BH merger candidate GW190814. We use the Bayesian framework presented by \cite{chen_2_2018} and \cite{gray_cosmological_2019}, which is detailed in the Appendix. We outline our methodology starting with a thorough account of our assumed priors and input parameters used in gwcosmo. Note that we adopt the 02-H0 branch to perform these calculations, as it is the most stable at the time of writing.\footnote{We fix a few small syntax errors that prevented the code from running. Also,  we add a few lines of code to gwcosmo.py and to the gwcosmo-single-posterior script to import NS-BH merger priors. The bin size of the ``dl" array in skymap.marginalized\_distance is changed from 200 to 50 for optimization reasons. The effect on our results is insignificant.}
The preparation and injection of the GLADE 2.0 galaxy catalog is discussed in \hyperref[sec:catalog]{Section 2.2}. The HEALPIX localization skymaps used for all of the calculations are described in \hyperref[sec:skymap]{Section 2.3}.
We create a baseline test of our assumptions by comparing to \cite{the_ligo_scientific_collaboration_gravitational-wave_2019} and \cite{fishbach_standard_2019} using GW170817. We then explore parameter space to present multiple H$_0$ calculations for GW190814. 

\subsection{Priors and Input Parameters}
\label{sec:priors}

Our analysis is carried out with both a uniform and ``flat log prior" on H$_0$ over a set of different intervals.We use the definition for the flat log prior $p(H_0) \propto H_0^{-1}$ \citep{the_ligo_scientific_collaboration_gravitational-wave_2019}. Below, we describe the options passed to the \textit{gwcosmo\_single\_posterior} script in the gwcosmo code.The italicized items are presented in Table 1.
\begin{enumerate}[leftmargin=*]
\item The {\it mass distribution} is chosen to be either (a) BNS-uniform, a  binary neutron star distribution over the interval [$1.0\,M_{\odot}$, $3.0\,M_{\odot}$], (b) BNS-Gaussian, a symmetric Gaussian distribution centered on $\mu=1.35\,M_{\odot}$ with $\sigma=0.15\,M_{\odot}$ \citep{kiziltan_neutron_2010}, and (c) NSBH-uniform, a uniform neutron-star--black-hole distribution with a uniform NS mass distribution over [$1\,M_{\odot}$, $3\,M_{\odot}$] and a power-law BH mass distribution over [$5\,M_{\odot}$, $40\,M_{\odot}$]. The uniform component follows $p(m_2) =$ constant, while the power-law component takes the form $p(m_1) \propto m_1^{-\alpha}$, with the power-law index $\alpha =  1.6$. Here, $m_2$ is the secondary (NS) mass and $m_1$ is the primary (BH) mass. 
\item The {\it power spectral density (PSD)} parameter is associated with the detector sensitivity during either the O1, O2, or O3 observing runs (we choose O2 for GW170817 and O3 for GW190814).
\item The {\it completeness} parameter is defined as the ratio of the number of galaxies in a chosen galaxy catalog to the true number of galaxies in the cosmological volume. We discuss this in more detail in Section 2.2. and in the Appendix. \cite{gray_cosmological_2019} study the effects of this parameter on the H$_0$ posterior extensively on simulated merger data in Sections III and IV.
\item {\it Galaxy weighting} may be set to either ``False" (equal weights) or have $B$-band luminosity-dependent weights $\omega_i \propto L_{B}^i$. A luminosity-dependent weighting scheme follows the assumption that BH and NS merger rates scale with star-formation rates \citep{fong_locations_2013}.
\item {\it Luminosity threshold} gives the minimum $B$-band luminosity considered for the calculation; this parameter will be explored in depth in \hyperref[sec:results]{Section 3}.
\end{enumerate}

\noindent
We hold the following parameters constant throughout every calculation. 
\begin{enumerate}[leftmargin=*]
    \item \textit{Linear cosmology} is set to ``False" because we include galaxies with redshift $z > 0.1$.
    \item \textit{Posterior samples} is set to ``False", given that initially, at the time of writing this paper, there had not yet been a data release from the LIGO-Virgo Collaboration (LVC) for event GW190814. After submission of this paper, the LVC released the full posterior samples for GW190814, allowing us (upon revision) to properly account for biases introduced by the approximation described below. Results for the latter analysis are found in Section 3.2. When neither a posterior sample nor an EM counterpart is used, the three-dimensional (3-D) skymap (see {Section 2.3}) is passed as the gravitational-wave data to the \textit{skymap.marginalized\_distance} function in gwcosmo, which is a Gaussian approximation to the GW's distance posterior. 
    \item \textit{Basic pdet} is set to ``False" allowing us to take into account redshifted mass, $M_z = M(1+z)$ \citep{chen_mass-redshift_2019}.
    \item The \textit{uncertainty} parameter is set to ``True," taking into account the Gaussian uncertainties in redshift for each galaxy (see \hyperref[sec:catalog]{Section 2.2}).
    \item The \textit{rate evolution} parameter is set to ``False," which describes a constant merger rate $R(z)$ as it appears in Equation 11 of \cite{the_ligo_scientific_collaboration_gravitational-wave_2019}. We assume the $\Lambda$CDM model ($\Omega_m = 0.308$, $\Omega_{\Lambda} = 0.692$). The effect of choosing a merger rate with a dependence on redshift is shown extensively by \citet{the_ligo_scientific_collaboration_gravitational-wave_2019}. 
    
\end{enumerate}
\subsection{Using the GLADE Galaxy Catalog}
\label{sec:catalog}

 We use the GLADE v2.3 galaxy catalog throughout, adopting the parameters $\phi^* = 1.6 \times 10^{-2}h^3$\,Mpc$^{-3}$, where $h=0.7$ and $\beta=-1.07$ for the Schechter $B$-band luminosity function,
\begin{equation}
\label{Eq.schechter}
\begin{aligned}
\rho(x)dx = \phi^*x^{\beta}e^{-x}dx, \ & \ &   x= L_{B}/L_{B}^*,
\end{aligned} 
\end{equation}
\noindent
where $\rho(x)$ is the number density of galaxies and the
characteristic $B$-band luminosity $L_{B}^*$ corresponds to an absolute magnitude $M_B = -20.47$ \citep{gehrels_galaxy_2016}.

The GLADE galaxy catalog contains nearly 3 million galaxies  and is complete up to 300\,Mpc at $L_B = 0.626\,L_B^*$, corresponding to the median of the luminosity function \citep{arcavi_optical_2017}.
The statistical method is sensitive to the galaxy completeness fraction, $f$ \citep{gray_cosmological_2019}. The Glade catalog's high completeness fraction over the redshifts considered for this study provides a significant advantage over other catalogs \citep{dalya_glade_2018}. Approximately half of the objects in the catalog have a measured $B$-band luminosity.

All redshifts in GLADE are corrected for peculiar motions and are in the heliocentric frame \citep{carrick_cosmological_2015}. In order to correct for radial group velocities, we cross-reference GLADE galaxies by their ``PGC ID'' with the Principal Galaxy Catalog (PGC) and use the corresponding corrected radial velocities in the heliocentric frame \citep{kourkchi_galaxy_2017}. We then correct these heliocentric velocities to the CMB frame using NASA/IPAC Extragalactic Database (NED) with parameters $l_{\rm apex} = 264.14^\circ$, $b_{\rm apex} = +48.26^\circ$, and  $v_{\rm apex} = 371.0$ km s$^{-1}$ \citep{fixsen_cosmic_1996}. We define the $z_{\text{rad}}$ and $z_{\text{CMB}}$ parameters as the radial group velocity and CMB reference frame corrections, respectively.  Finally, we assign a 200 km s$^{-1}$ Gaussian uncertainty to the velocity $(cz)$ for each galaxy.

For our purposes, we only extract the right ascension (degrees), declination (degrees), redshift, apparent $B$ magnitude, and absolute $B$ magnitude (RA, Dec, $z$, $B$, $B_{\rm abs}$, respectively) from the raw catalog and build the pickle\footnote{https://docs.python.org/3/library/pickle.html .}-formatted dictionary that gwcosmo requires. Note that gwcosmo expects the RA and Dec to be in radians. We use the provided $B$-band magnitude in our 
galaxy-weighting procedures.

\subsection{Skymap Localization}
\label{sec:skymap}

\begin{figure}
    \centering
    \includegraphics[width=0.45\textwidth]{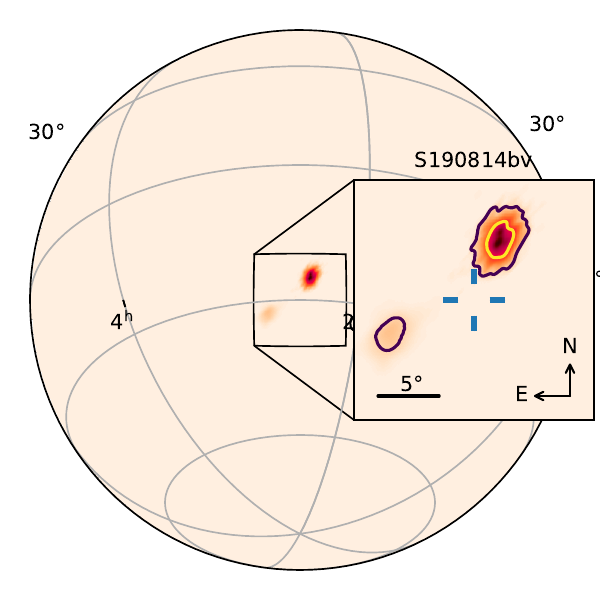}
    \caption{Globe projected skymap for GW190814 (S190814bv) using the \textit{ligo-skymap-plot} module. The inset shows a close-up view centered on (RA = 1$^{\rm hr}$, Dec = $-30^\circ$; blue tick marks). Dark purple and yellow contours represent the 90\% (23 deg$^2$) and 50\% (5 deg$^2$) credible regions, respectively.}
    \label{fig:my_label}
\end{figure}

\begin{figure*}
    \centering
    \includegraphics[width=0.9\textwidth]{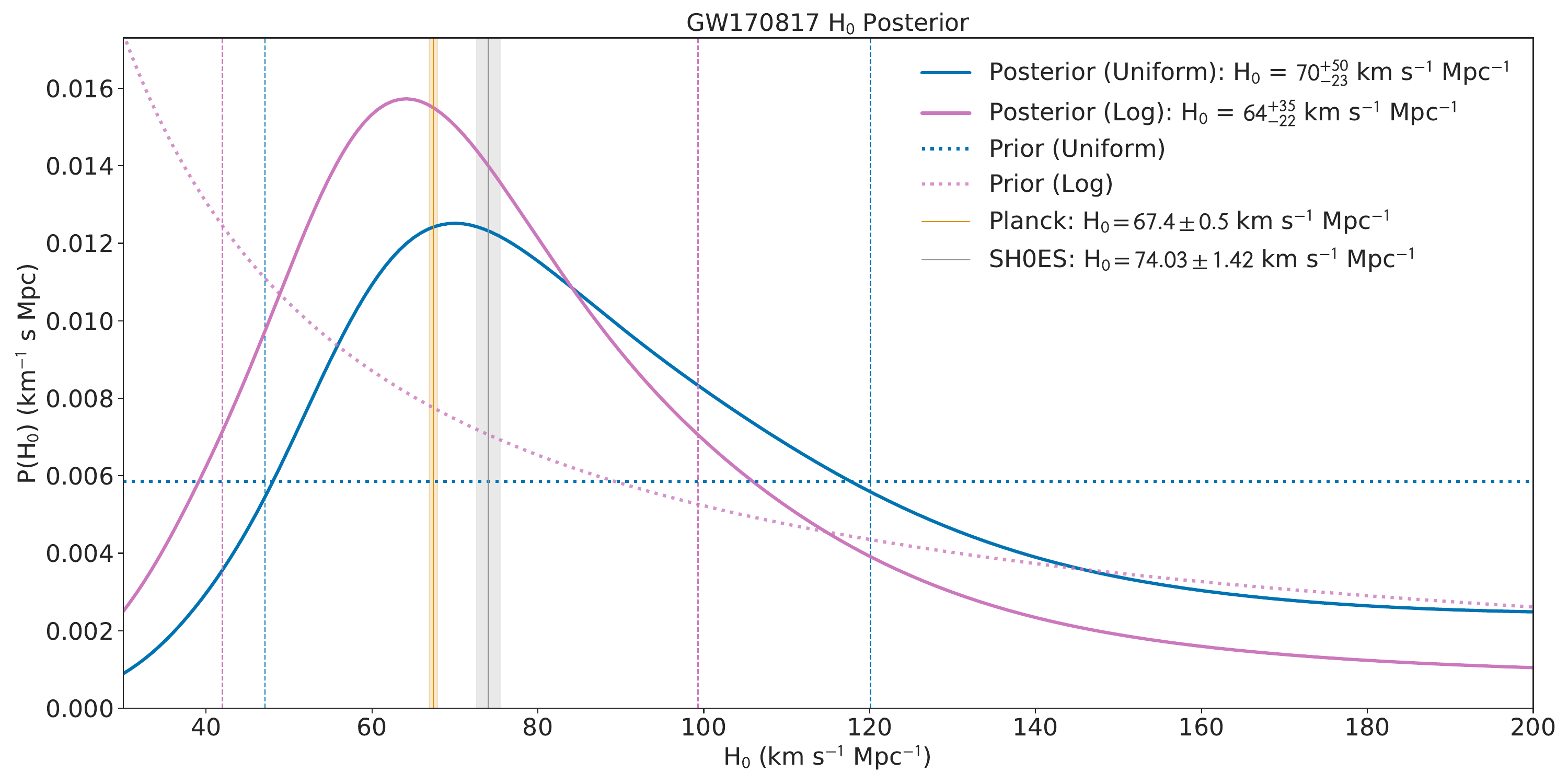}
    \caption{H$_0$ posterior for GW170817 (no-counterpart) assuming a flat log (purple) and uniform (blue) prior on H$_0$ along the interval [30,~200]\,km\,s$^{-1}$\,Mpc$^{-1}$. Here, we adopt the Gaussian BNS mass distribution with $\mu = 1.35\,M_{\odot}$, $\sigma = 0.15\,M_{\odot}$. The $B$-band luminosity threshold is $L_B \geq 0.001\,L_B^*$. All galaxies have equal luminosity weights and negligible redshift uncertainties. We include the most recent {\it Planck} (orange vertical line) and SH0ES (grey vertical line) measurements and their corresponding 1$\sigma$ uncertainties (shaded regions).}
    \label{fig:my_label}
\end{figure*}

\begin{deluxetable*}
{lccccccccccc}
\tablecaption{Input parameters and corresponding H$_0$ posterior results for GW190814 \label{tab:param}}
\tablewidth{0pt}
\tablehead{
\colhead{Row} & \colhead{Luminosity Threshold} & \colhead{Redshift Correction} & \colhead{Galaxy Weighting} & \colhead{H$_0$ Posterior} \\
\colhead{Name} & \colhead{$B$ band} & \colhead{Type} & \colhead{Bool} & \colhead{68\% (km\,s$^{-1}$\,Mpc$^{-1}$)}
}
\startdata
A & $\geq 0.001\,L_B^*$ & z$_{\text{rad}}$, z$_{\text{CMB}}$ & False & $67^{+41.0}_{-26.0}$ \\
B & $\geq 0.01\,L_B^*$ & z$_{\text{rad}}$, z$_{\text{CMB}}$ & False & $68^{+39.0}_{-28.0}$  \\
C & $\geq 0.001\,L_B^*$ & z$_{\text{rad}}$, z$_{\text{CMB}}$ & True & $66^{+55.0}_{-12.0}$ \\
D & $\geq 0.25\,L_B^*$ & z$_{\text{rad}}$, z$_{\text{CMB}}$ & False & $69^{+39.0}_{-28.0}$ \\
E & $\geq 0.626\,L_B^*$ & z$_{\text{rad}}$, z$_{\text{CMB}}$ & False & $71^{+34.0}_{-30.0}$ \\
F & $\geq 0.001\,L_B^*$ & None & False & $68^{+40.0}_{-27.0}$ \\
G & $\geq 0.01\,L_B^*$ & None & False & $68^{+40.0}_{-27.0}$  \\
H & $\geq 0.001\,L_B^*$ & None & True & $66^{+52.0}_{-15.0}$ \\
I & $\geq 0.25\,L_B^*$ & None & False & $70^{+48.0}_{-19.0}$ \\
J & $\geq 0.626\,L_B^*$ & None & False & $72^{+42.0}_{-23.0}$ \\
\enddata
\tablecomments{The z$_{\text{rad}}$ and z$_{\text{CMB}}$ labels signify that we applied the radial group velocity and CMB reference frame corrections discussed in \hyperref[sec:catalog]{Section 2.2}. The following parameters are constant for each measurement. \textit{H$_0$ prior}, Uniform [40,~140] (km\,s$^{-1}$\,Mpc$^{-1}$); \textit{PSD}, O3; \textit{mass distribution}, NSBH-Uniform; \textit{completeness}, False.  }
\end{deluxetable*}

\begin{figure*}
    \centering
    \includegraphics[width=1.0\textwidth]{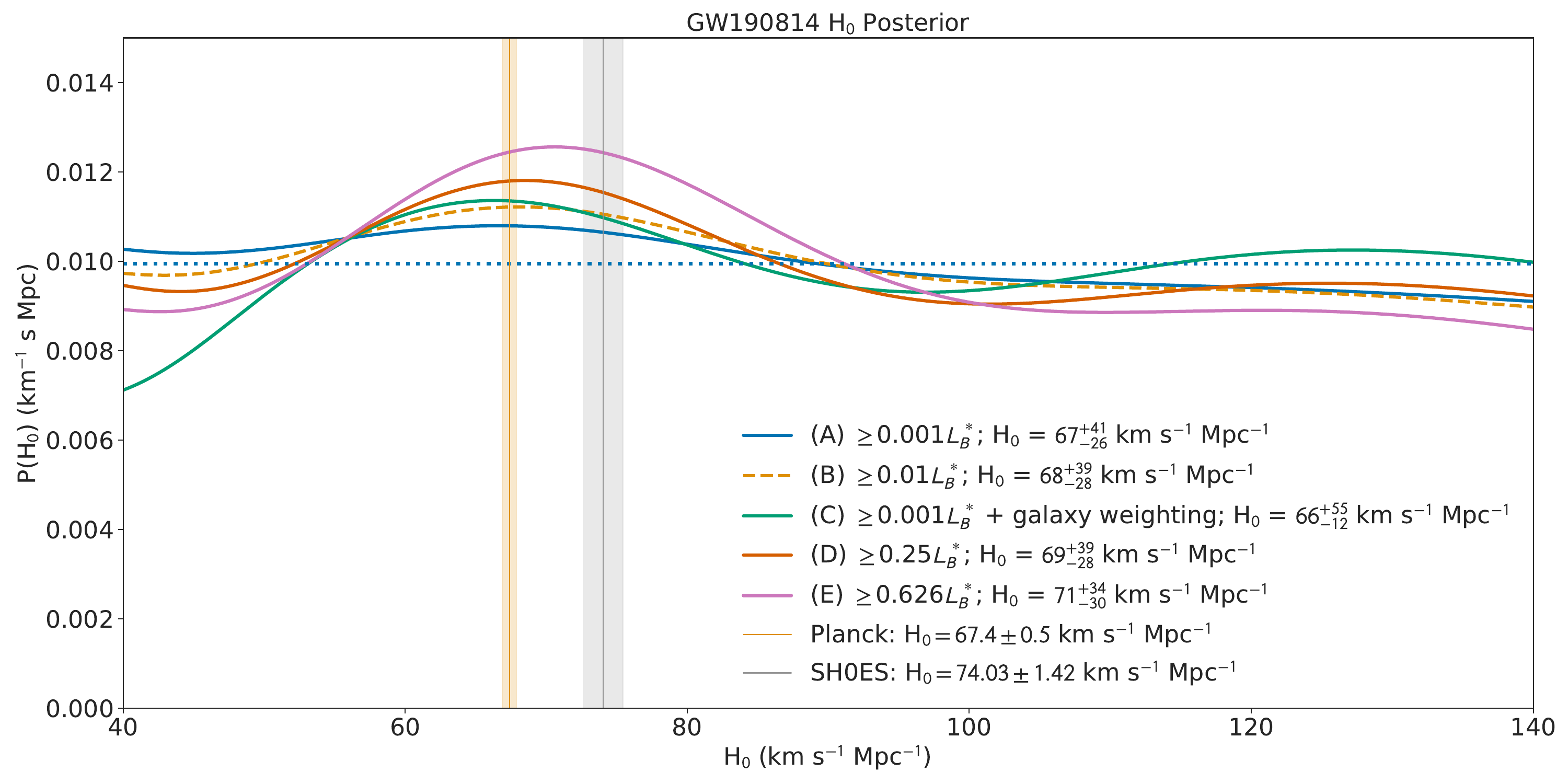}
    \caption{Combined H$_0$ posterior for GW190814 assuming a uniform (dotted blue) H$_0$ prior along the interval [40,~140]\,km\,s$^{-1}$\,Mpc$^{-1}$ . Lettering in the legend corresponds to the rows in Table 1. We illustrate the incremental appearance of structure in the posterior as the luminosity threshold is increased. For A (solid blue), B (dotted orange), D (solid red orange), and E (pink), we set a luminosity threshold of $0.001\,L_B^*$, $0.01\,L_B^*$, $0.25\,L_B^*$, and $0.626\,L_B^*$, respectively. C (solid green) shows the posterior when the luminosity threshold is $0.001\,L_B^*$ and \textit{galaxy weighting} is set to ``True." We include the most recent {\it Planck} (orange vertical line) and SH0ES (gray vertical line) measurements and their corresponding 1$\sigma$ uncertainties (shaded regions).}
    \label{fig:my_label}
\end{figure*}

During O2 and O3, LIGO-Virgo released public alerts accompanied by an allsky HEALPix localization skymap in a FITS file format for each GW event.
The skymap FITS includes the sky position, probability, and
distance for each pixel. The original 90\% region for GW170817
(28 deg$^2$)  was improved to 16 deg$^2$ using the LALInference pipeline \citep{veitch_robust_2015}. For our analysis, we adopt the
updated skymap.\footnote{We use the updated GW170817 skymap
named \href{https://dcc.ligo.org/LIGO-P1800061/public}{figure\_3.tar.gz} .}

For GW190814, we use the updated skymap from the GraceDB database, localizing the source to 23 deg$^2$ and 5 deg$^2$ in the 90\% and 50\% confidence regions, respectively\footnote{The publicly available updated S190814\small{bv} skymap LALInference.v1.fits.gz is available at \href{https://gracedb.ligo.org/superevents/GW190814/view/}{GraceDb}.}. The marginalized distance posterior found in the skymap is a symmetric Gaussian fit to the full, potentially asymmetric posterior sample. According to \cite{fishbach_standard_2019}, this assumption has the effect of moving the peak of the H$_0$ posterior by as much as 9\% compared to using a full posterior sample. For our calculations, Equation A5 of the Appendix only includes galaxies in the 99.9\% region (assigning weight = 1 to each if equal weights) and assigns a weight of 0 for those outside of the localization region. 

\section{Results}
\label{sec:results}
We split our analysis into two parts. First, we summarize our results for GW170817 and compare our H$_0$ inference to previous works using this statistical method \citep{gray_cosmological_2019,fishbach_standard_2019,the_ligo_scientific_collaboration_gravitational-wave_2019}. We also explore possible systematic differences and assumptions that may carry into our H$_0$ calculation for GW190814. We then repeat the procedure over the parameter space introduced in Section 2 for GW190814. All H$_0$ measurements assume the 68.3\% highest density posterior interval. We choose the H$_0$ prior range as [30,~200] or [40,~140]\,km\,s$^{-1}$\,Mpc$^{-1}$ for ease of comparison to other works. Finally, we repeat the calculation using the full posterior samples released by the LVC on June 25, 2020.

We note that after submission and initial review of this paper, the Dark Energy Survey (DES) Collaboration posted a paper  \citep{palmese_statistical_2020} in which a similar analysis of GW190814 is presented, but they did not include posterior samples.

\subsection{Statistical Method}

Since the GLADE catalog is 100\% complete up to $z=0.03$ for galaxies that are above  $0.25\,L_B^*$, and we consider galaxies well above this redshift threshold and down to $0.001\,L_B^*$ ($M_B = -12.96$\,mag), we set the completeness parameter to ``False" in our final calculation unless stated otherwise. We consider a flat log and uniform prior on H$_0$ on the intervals [30,~200] and [40,~140]\,km\,s$^{-1}$\,Mpc$^{-1}$. We use the BNS-Gaussian mass distribution centered on $\mu=1.35\,M_{\odot}$ and a standard deviation of $\sigma=0.15\,M_{\odot}$. Galaxies up to $z \approx 0.5$ ($z_{\rm max} = 0.5$) are allowed, with equal luminosity weights assigned. A 200 km s$^{-1}$ Gaussian uncertainty is applied to the velocity $(cz)$ of each galaxy. 2662 galaxies fall into the 99\% skymap localization corresponding to 34 deg$^2$. We employ a constant rate evolution term discussed in Section 2.1. The resulting H$_0$ posterior is illustrated in Figure 2. 

Given a uniform H$_0$ prior along the intervals [30,~200] and [40,~140]\,km\,s$^{-1}$\,Mpc$^{-1}$, we infer a Hubble constant $70^{+50.0}_{-23.0}$\,km\,s$^{-1}$\,Mpc$^{-1}$ and $70^{+35.0}_{-18.0}$\,km\,s$^{-1}$\,Mpc$^{-1}$, respectively. Even when accounting for a luminosity cut $\geq 0.626\,L_B^*$ yielding H$_0 = 73^{+36.0}_{-17.0}$, our peak is less pronounced and is shifted compared to the H$_0 \geq 74$\,km\,s$^{-1}$\,Mpc$^{-1}$ obtained by \cite{fishbach_standard_2019}. This discrepancy may be caused by differences in both our chosen subset of the GLADE galaxy catalog and our velocity corrections. As a qualitative test, we also calculate H$_0$ without the radial group velocity and CMB reference frame corrections, yielding $67^{+37.0}_{-19.0}$ km s$^{-1}$ Mpc$^{-1}$ for a uniform H$_0$ prior interval, [40,~140]\,km\,s$^{-1}$\,Mpc$^{-1}$. Given the sensitivity of the posterior to the injected catalog, we expect a slight deviation if the catalog is handled differently. These systematic differences will carry over into our calculation for the NS-BH merger.

For GW190814, we assume the same $B$-band Schechter parameters chosen for the GW170817 calculations. Here, we focus only on the [40,~140]\,km\,s$^{-1}$\,Mpc$^{-1}$  prior range on $\text{H}_0$. We show our results in Table 1 and illustrate the H$_0$ posterior in Figure 3, where we plot five realizations for different luminosity considerations. We apply a maximum redshift limit $z_{\rm max} = 0.5$ and assume the NSBH-uniform mass distribution. 64,735 galaxies are obtained in the 99\% localization region.
We repeat the calculation for GW190814 to obtain H$_0 =  67^{+41.0}_{-26.0}$\,km\,s$^{-1}$\,Mpc$^{-1}$ and H$_0 = 71^{+34.0}_{-30.0}$\,km\,s$^{-1}$\,Mpc$^{-1}$ for  $L_B \geq 0.001\,L_B^*$ and $L_B \geq 0.626\,L_B^*$, respectively. Our tested parameter space is detailed in Table 1. According to Figure 3, the posterior peak is more pronounced for stricter luminosity cuts. The peak of our H$_0$ posterior shifts by $\sim 6$\% over the range of luminosity cuts, which is in agreement with \cite{fishbach_standard_2019}.
The GW190814 posterior appears significantly flatter compared to GW170817. Given that the GW190814 localization covered a larger volume and included more than ten times as many galaxies as GW170817, we expect the GW190814 posterior to be washed out. The bump in the H$_0 = 100$--140\,km\,s$^{-1}$\,Mpc$^{-1}$ range is enhanced when \textit{galaxy weighting} is set to ``True"; it may be an artifact of enhanced GLADE catalog features similar to Figure 2 of  \citet{the_ligo_scientific_collaboration_gravitational-wave_2019}.

\begin{figure*}
    \centering
    \includegraphics[width=1.0\textwidth]{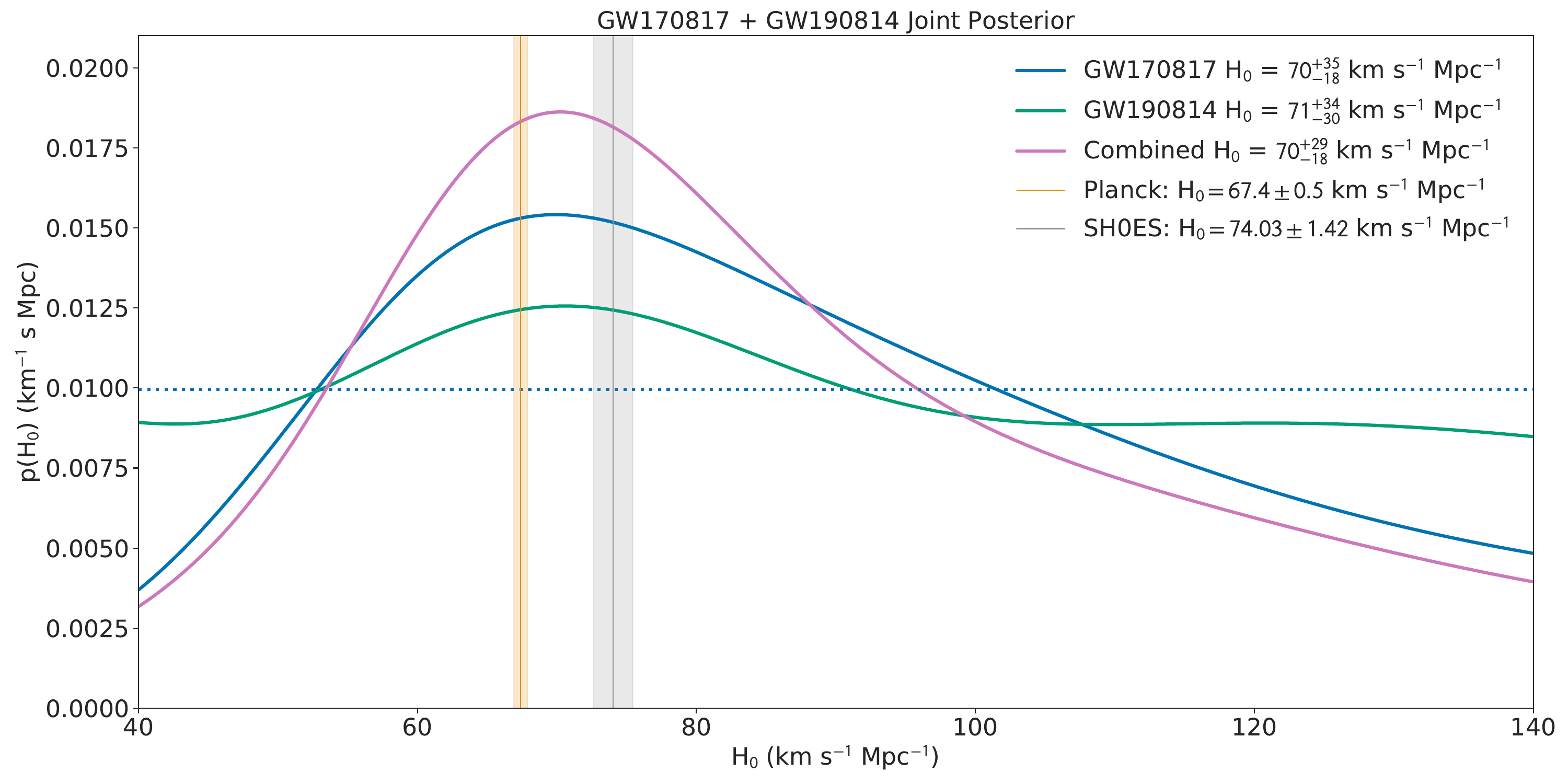}
    \caption{Combined H$_0$ posterior for GW170817 and GW190814 assuming uniform (dotted blue) H$_0$ prior along the interval [40,140]\,km\,s$^{-1}$\,Mpc$^{-1}$. For GW190814, we use the parameters in Row E from Table 1. We include the most recent {\it Planck} (orange vertical line) and SH0ES (gray vertical line) measurements and their corresponding 1$\sigma$ uncertainties (shaded regions).}
    \label{fig:my_label}
\end{figure*}

We then combine the posteriors for GW170817 with S190814 using the \textit{gwcosmo\_combined\_posterior} script, yielding  H$_0 = 70^{+29.0}_{-18.0}$\,km\,s$^{-1}$\,Mpc$^{-1}$ as shown in Figure 4. Although the peak is centered between the {\it Planck} and SH0ES results, we caution that the value is subject to systematics arising from luminosity cuts or weighting. For example, our combined posterior used a luminosity threshold of $0.626\,L_B^*$ for GW190814, whereas taking the luminosity down to $0.001\,L_B$ would produce a peak at H$_0 \leq 70$\,km\,s$^{-1}$\,Mpc$^{-1}$. The combined posterior demonstrates the ability to further constrain the Hubble constant with multiple GW sources.

Systematic biases in the joint posterior due to varying population parameters of astrophysical sources are expected to be smaller than the statistical uncertainties due to contributions from the galaxy catalog given a high probability that the host galaxy is in the catalog. GW190814 has a median source redshift of $z_{\rm event} = 0.053$ using the combined waveform model from \citet{abbott_gw190814_2020}, corresponding to $p(G|z_{\rm event},D_w) > 0.6$ (or ``high in-catalog probability") with the GLADE catalog according to Figure 1 of \citet{the_ligo_scientific_collaboration_gravitational-wave_2019}. Therefore, we take the contributions from the galaxy catalog to be the dominant source of uncertainties for GW190814. \\
\\
\\

\subsection{Using the Posterior Samples}
\begin{figure*}
    \centering
    \includegraphics[width=1.0\textwidth]{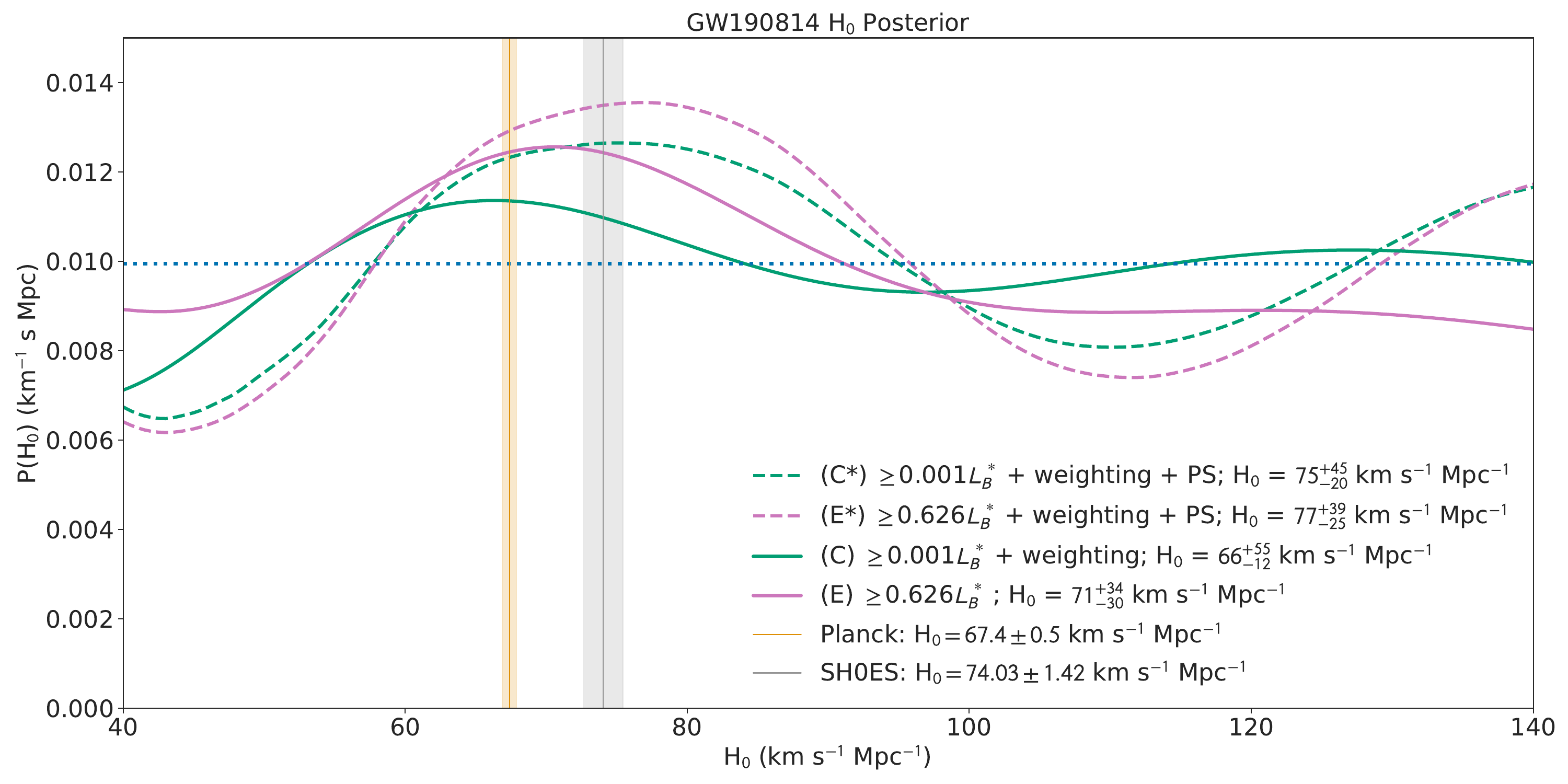}
    \caption{H$_0$ posterior for GW190814 assuming a uniform (dotted blue) H$_0$ prior along the interval [40,~140]\,km\,s$^{-1}$\,Mpc$^{-1}$.
     Lettering in the legend corresponds to the rows in Table 1. We illustrate the difference between using posterior samples (a combined SEOBNRv4PHM and IMRPhenomPv3HM waveform model) and using a Gaussian approximation to the distance posterior from the 3D skymap. The C (solid green) and E (solid pink) curves are identical to those in Figure 4. We label C* (dashed green) and E* (dashed purple) to indicate when posterior samples are used. We include the most recent {\it Planck} (orange vertical line) and SH0ES (gray vertical line) measurements and their corresponding 1$\sigma$ uncertainties (shaded regions).}
    \label{fig:my_label}
\end{figure*}

In light of the LVC data release for GW190814 on June 25, 2020, we apply the full posterior sample to our H$_0$ calculation \citep{abbott_gw190814_2020}. We now account for biases introduced when a Gaussian approximation to the distance posterior is used via the 3D skymap. For GW170817, we use the low-spin PhenomPNRT posterior sample\footnote{The posterior sample file can be found at \href{https://dcc.ligo.org/LIGO-P1800061/public}{LIGO-P1800061-v11.}}. For GW190814, we use a combined posterior sample consisting of the SEOBNRv4PHM (EOBNR PHM; \citealt{babak_validating_2017}; \citealt{ossokine_multipolar_2020}) and IMRPhenomPv3HM (Phenom PHM; \citealt{khan_phenomenological_2019,khan_including_2020}) Waveform Models.\footnote{Relevant files can be accessed at \href{https://www.gw-openscience.org/eventapi/html/O3_Discovery_Papers/GW190814/v1/}{Gravitational Wave Open Science Center.}} In Figure 6, we compare the differences in the GW190814 H$_0$ posterior with and without the use of posterior samples.

Following the reasoning of \citet{fishbach_standard_2019}, the use of a full posterior sample (accounting for masses and spins) as opposed to a Gaussian approximation to the distance posterior can have the effect of shifting the H$_0$ posterior peak. In our case, we observe a notable shift for both sets of parameters, favoring a higher value for the Hubble constant when using posterior samples. In Figure 6, we show the combined H$_0$ posterior following the same procedure used to produce Figure 4, but now with posterior samples for both GW170817 and GW190814. We obtain H$_0 = 67^{+36.0}_{-15.0}$\,km\,s$^{-1}$\,Mpc$^{-1}$ (GW170817; $L_B \geq 0.001\,L_B^*$), H$_0 = 75^{+45.0}_{-20.0}$\,km\,s$^{-1}$\,Mpc$^{-1}$ (GW190814; $L_B \geq 0.001\,L_B^*$), and H$_0 = 77^{+39.0}_{-25.0}$\,km\,s$^{-1}$\,Mpc$^{-1}$ (GW190814; $L_B \geq 0.626\,L_B^*$), in agreement with the results of \cite{abbott_gw190814_2020}. Our combined H$_0$ calculation with posterior samples yields H$_0 = 69^{+29.0}_{-14.0}$\,km\,s$^{-1}$\,Mpc$^{-1}$. Here, we observe a slight improvement to the uncertainty in the combined H$_0$ inference, owing to the additional information gained when using posterior samples.

\begin{figure*}
    \centering
    \includegraphics[width=1.0\textwidth]{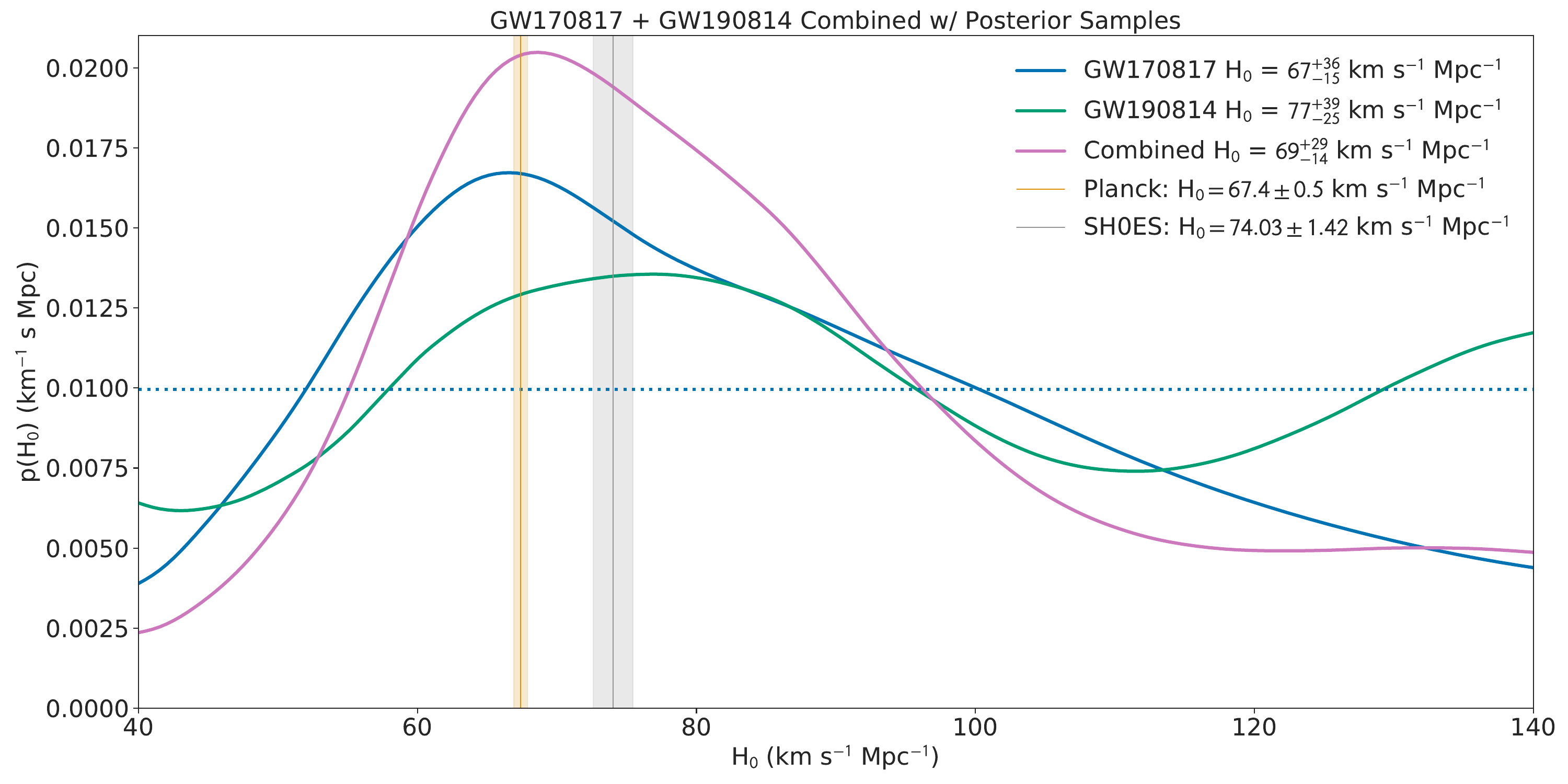}
    \caption{Combined H$_0$ posterior for GW170817 and GW190814 assuming uniform (dotted blue) H$_0$ prior along the interval [40,~140]\,km\,s$^{-1}$\,Mpc$^{-1}$. The GW170817 posterior (solid blue line) is now calculated using a low spin PhenomPNRT posterior sample. For GW190814, we use the parameters in Row E from Table 1 together with a SEOBNRv4PHM + IMRPhenomPv3HM posterior sample (labeled E* in Figure 5). We include the most recent {\it Planck} (orange vertical line) and SH0ES (gray vertical line) measurements and their corresponding 1$\sigma$ uncertainties (shaded regions).}
    \label{fig:my_label}
\end{figure*}

\section{Conclusions}\label{s:conc}
We demonstrated the statistical catalog method to infer the value of H$_0$ for GW sources without an optical counterpart. Previous measurements have been made for GW events in O1 and O2, most notably with GW170817. Our baseline test served as a calibration against the result of \cite{fishbach_standard_2019} to identify systematic differences that may affect our H$_0$ measurement for NS-BH merger GW190814. Using a uniform H$_0$ prior [40,140]\,km\,s$^{-1}$\,Mpc$^{-1}$ with a luminosity threshold of $0.001\,L_B^*$ for both GW170817 and GW190814, we infer H$_0 = 70^{+35.0}_{-18.0}$\,km\,s$^{-1}$\,Mpc$^{-1}$ and $67^{+41.0}_{-26.0}$\,km\,s$^{-1}$\,Mpc$^{-1}$, respectively. We then increased the luminosity threshold up to $L_B \geq 0.626\,L_B^*$ and obtain H$_0 = 71^{+34.0}_{-30.0}$\,km\,s$^{-1}$\,Mpc$^{-1}$ for GW190814. This tighter value is used in combination with the GW170817 posterior to achieve a final value of H$_0 = 70^{+29.0}_{-18.0}$\,km\,s$^{-1}$\,Mpc$^{-1}$. We repeat the individual and joint inferences using the low-spin PhenomPNRT and combined (SEOBNRv4PHM + IMRPhenomPv3HM) posterior samples for GW170817 and GW190814, respectively. We achieve a tighter constraint for the joint measurement with H$_0 = 69^{+29.0}_{-14.0}$\,km\,s$^{-1}$\,Mpc$^{-1}$ when using posterior samples. Several sources of systematics were identified, including the injected catalog, luminosity weighting, and luminosity thresholds. 

The motivation for this method  follows the expectation of having significantly more GW-only mergers (dark sirens) than mergers with optical counterparts, providing a valuable test for H$_0$ inferences using the electromagnetic counterpart. As we head into the next generation of gravitational-wave detectors, the standard-siren method will improve the constraints on the Hubble constant. This new independent method may potentially resolve the Hubble tension problem or compel us to reevaluate our cosmological models, specifically $\Lambda$-CDM.

\begin{appendix}
\renewcommand{\theequation}{A\arabic{equation}}
\begin{large}
\begin{center}
{\it The Statistical Method}
\end{center}
\end{large}
Here, we summarize the statistical method adopted from \cite{chen_2_2018}, \cite{gray_cosmological_2019}, and \cite{the_ligo_scientific_collaboration_gravitational-wave_2019}. For relevant formalism, also see \citet{mandel_extracting_2019}, \citet{thrane_introduction_2019}, and \citet{vitale_one_2020}.

The posterior probability on H$_0$ from $N$ gravitational-wave (GW) events can be computed as
\begin{equation}
\begin{aligned}
p(\text{H}_0|x_{\text{GW}},D_{\text{GW}})&=\frac{p(\text{H}_0)p(N|\text{H}_0)\prod_i^{N} p({x_{\text{GW}}}_i|{D_{\text{GW}}}_i,\text{H}_0)}{p(x_{\text{GW}}|D_{\text{GW}})}
\\ &\propto p(\text{H}_0)\prod_i^{N} p({x_{\text{GW}}}_i|{D_{\text{GW}}}_i,\text{H}_0),
\end{aligned}
\end{equation}
where $x_{\text{GW}}$ is the set of GW data and $D_{\text{GW}}$ indicates that the detection was made in the form of a GW.  Here, $p({\text{H}}_0)$ is the prior on H$_0$ that we took to be either uniform or flat log over an interval [a,b]. $p(N|\text{H}_0)$ is the likelihood of detecting $N$ events given a value for $\text{H}_0$. Using the same prior on the astrophysical rate of events as in \cite{the_ligo_scientific_collaboration_gravitational-wave_2019}, we drop the dependence of this term on $\text{H}_0.$
\\

For an individual gravitational-wave event, the likelihood can be written as
\begin{equation}
\label{Eq.xD}
\begin{aligned}
p(x_{\text{GW}}|D_{\text{GW}},\text{H}_0) &= \dfrac{p(D_{\text{GW}}|x_{\text{GW}},\text{H}_0)p(x_{\text{GW}}|\text{H}_0)}{p(D_{\text{GW}}|\text{H}_0)},
\\ &= \dfrac{p(x_{\text{GW}}|\text{H}_0)}{p(D_{\text{GW}}|\text{H}_0)},
\end{aligned} 
\end{equation}
\\

The normalization factor $p(D_{\text{GW}}|\text{H}_0)$ in the denominator of Eq.~\ref{Eq.xD} can be evaluated with the integral
\begin{equation}
\label{Eq.pdet}
\begin{aligned}
p(D_{\text{GW}}|\text{H}_0) &= \int p(D_{\text{GW}}|x_{\text{GW}},\text{H}_0)
p(x_{\text{GW}}|\text{H}_0) dx_{\text{GW}}
\\ &= \int^{\infty}_{\rho > \rho_{\rm th}}p(x_{\text{GW}}|\text{H}_0) dx_{\text{GW}},
\end{aligned}
\end{equation}
where $\rho_{\rm th}$ is the signal-to-noise ratio (SNR) threshold below which $p(D_{\text{GW}}|\text{H}_0) = 0$. In our calculations we assume $\rho_{\rm th} = 8$, the default in gwcosmo.
Calculating the event detectability, $p(D_{\text{GW}}|H_{0})$, involves marginalizing over masses, inclination, polarization, and sky location. Gwcosmo (detection\_probability.py) uses a Monte-Carlo integration to marginalize over many GW events. The source masses are drawn from the prior mass distribution $p(m_1,m_2)$, defined in Section 2.1. The source masses are converted to observed masses when setting \textit{Basic pdet} to ``False". Treatment of mass distributions is described in more detail in Appendix 5 of \cite{gray_cosmological_2019}.\footnote{The calculation of Eq. (A3) has been a widely explored topic in the literature. Systematic biases can arise when this quantity does not fully account for all of the parameters on which the gravitational-wave signal is dependent. We emphasize that these effects are insignificant compared to the statistical contribution from the galaxy catalog, as discussed in Section 3.1.}
\\

In our case, we use the galaxy-catalog method, in which the likelihood function $p(x_{\text{GW}}|D_{\text{GW}},\text{H}_0)$ can be expanded as
\begin{equation} \label{Eq:sum G}
\begin{aligned}
p(x_{\text{GW}}|D_{\text{GW}},\text{H}_0) &= \sum_{g=G,\bar{G}} p(x_{\text{GW}},g|D_{\text{GW}},\text{H}_0)
\\ &= \sum_{g=G,\bar{G}} p(x_{\text{GW}}|g,D_{\text{GW}},\text{H}_0) p(g|D_{\text{GW}},\text{H}_0)
\\ &= p(x_{\text{GW}}|G,D_{\text{GW}},\text{H}_0) p(G|D_{\text{GW}},\text{H}_0) + p(x_{\text{GW}}|\bar{G},D_{\text{GW}},\text{H}_0) p(\bar{G}|D_{\text{GW}},\text{H}_0).
\end{aligned} 
\end{equation}
with $G$ and $\bar{G}$ denoting the cases where the host is in the catalog and where it is not, respectively.\\

The likelihood \textit{when} the host
galaxy is in the catalog, $p( x_{\text{GW}}|G, D_{\text{GW}}, \text{H}_0)$, can be written as
\begin{equation} \label{Eq:p(x|G,D,H0)}
\begin{aligned}
p(& x_{\text{GW}}|G, D_{\text{GW}}, \text{H}_0) = \dfrac{\sum^{N_\text{gal}}_{i=1} \int p(x_{\text{GW}}|z_i,\Omega_i,\text{H}_0)p(s|M(z_i,m_i,\text{H}_0)) p(z_i) dz_i}
{\sum^{N_\text{gal}}_{i=1} \int p(D_{\text{GW}}|z_i,\Omega_i,\text{H}_0)p(s|M(z_i,m_i,\text{H}_0)) p(z_i) dz_i} ,
\end{aligned}
\end{equation}
where $N_{\rm gal}$ is the number of galaxies considered in the catalog, $\Omega(\alpha,\delta)$ is the sky position angle, $z_i$ is the redshift of galaxy, $s$ indicates that a GW has been emitted (distinct from detected), and $m_i$ and $M$ are respectively the apparent and absolute magnitude of the galaxy.  
\\ 

Luminosity weighting:
\begin{equation}
\begin{aligned}
p(s|M,\text{H}_0) &\propto 
\begin{cases}
L(M(\text{H}_0)) & \text{galaxy (luminosity) weighting = True}\\
\text{const.} & \text{galaxy weighting = False.}
\end{cases}
\end{aligned}
\end{equation}
\\

Redshift evolution rate:
\begin{equation}
\begin{aligned}
p(s|z) &\propto 
\begin{cases}
(1+z)^{\lambda} & \text{if rate evolves with redshift}\\
\text{const.} & \text{if rate is constant with redshift,}
\end{cases}
\end{aligned}
\end{equation}
\\
where $\lambda$ is the rate evolution parameter, whose default value is 3.0 in gwcosmo. However, we hold the rate constant throughout all calculations. We define $V_c(z)$ as the co-moving volume contained within a redshift $z$.
\\

Redshift prior:
\begin{equation}
\begin{aligned}
p(z)&\propto \frac{1}{1+z}\frac{V_c(z)}{dz},  \text{if merger rate density = const.}
\end{aligned}
\end{equation}
\\
\noindent
The likelihood when the host galaxy is not in the catalog is defined as
\begin{equation}
\begin{aligned}
p(x_{\text{GW}}|\bar{G},D_{\text{GW}},\text{H}_0) &= \dfrac{\int  \int \int^\infty_{z(M,m_{\text{th}},\text{H}_0)}  p(x_{\text{GW}}|z,\Omega,\text{H}_0) p(z)p(\Omega)p(s|M,\text{H}_0)p(M|\text{H}_0)dz  d\Omega dM}{\int \int \int^\infty_{z(M,m_{\text{th}},\text{H}_0)}  p(D_{\text{GW}}|z,\Omega,\text{H}_0) p(z)p(\Omega)p(s|M,\text{H}_0)p(M|\text{H}_0)dz  d\Omega dM}.
\end{aligned}
\end{equation}

\begin{equation}
\label{Eq:G_DH0_end}
\begin{aligned}
\\ p(G|D_{\text{GW}},\text{H}_0)&= \dfrac{\int \int \int^{z(M,m_{\text{th}},\text{H}_0)}_0 p(D_{\text{GW}}|z,\Omega,\text{H}_0) p(s|z)p(z)p(\Omega)p(s|M,\text{H}_0)p(M|\text{H}_0) dz d\Omega dM}{\int \int \int^{\infty}_0 p(D_{\text{GW}}|z,\Omega,\text{H}_0) p(s|z)p(z)p(\Omega)p(s|M,\text{H}_0)p(M|\text{H}_0) dz d\Omega dM},
\end{aligned}
\end{equation}

\begin{equation}
\label{Eq:Gbard}
\begin{aligned}
p(\bar{G}|D_{\text{GW}},\text{H}_0)&= 1-p(\bar{G}|D_{\text{GW}},\text{H}_0).
\end{aligned}
\end{equation}
Equations \ref{Eq:G_DH0_end} and \ref{Eq:Gbard} are the probabilities that the host {\it is} and {\it is not} in the galaxy catalog, respectively.  
\\
\noindent
The prior on the GW host-galaxy sky location, $p(\Omega)$, is taken to be uniform across the sky. 
\\
\noindent
The prior on the absolute magnitude, $p(M|\text{H}_0)$, is taken to be proportional to the \citet{schechter_analytic_1976} luminosity function, the parameters for which are defined in \hyperref[sec:catalog]{Section 2.2}. Here, $m_{th}$ is the apparent magnitude threshold of the flux-limited galaxy catalog.

A complete description of the mathematics in gwcosmo is given in the Appendix of \cite{gray_cosmological_2019}.

\end{appendix}

\begin{acknowledgments} 

We thank Benjamin Stahl and Keto Zhang (U.C. Berkeley) for helpful advice on writing this paper, as well as the anonymous referee whose suggestions improved its quality. Ignacio Maga$\tilde{\text{n}}$a Hernandez (U.W. Milwaukee) discussed gwcosmo with us. We are also grateful to the lscsoft team for their efforts in providing public access to the gwcosmo code.
Generous financial support for this work was provided by Steven Nelson, the Christopher R. Redlich Fund, and the Miller Institute for Basic Research in Science (U.C. Berkeley). 

\bigskip

\end{acknowledgments}

\newpage
\bibliographystyle{aasjournal}
\bibliography{S190814bv}

\end{document}